# Spontaneous Synchrony Breaking

Research on synchronization of coupled oscillators has helped explain how uniform behavior emerges in populations of non-uniform systems. But explaining how uniform populations engage in sustainable non-uniform synchronization can prove to be just as fascinating.

Adilson E. Motter
Department of Physics and Astronomy
and Northwestern Institute on Complex Systems,
Northwestern University, Evanston, Illinois 60208, USA

Synchronization phenomena continue to inform and surprise. In classical and quantum physics, the harmonic oscillator is a paradigmatic model both because it describes periodic behavior found in a wide variety of systems and because it provides insights into more general behavior. Similarly, the study of networks of coupled oscillatory dynamical entities has a lot to offer for the understanding of emergent behavior in complex systems. The strongest form of such collective behavior is spontaneous synchronization, in which the oscillators coordinate their dynamics in a decentralized way. Spontaneous synchronization occurs in diverse contexts, from communities of chirping crickets to arrays of pulsing lasers. Its study has helped explain how non-identical entities, such as the crickets, can adjust their "rhythms" exclusively due to interactions. Writing in Physical Review Letters, Martens, Laing and Strogatz address the reciprocal to this question, namely a surprising scenario in which identical oscillators with identical coupling patterns self-organize into subpopulations with different synchronous behavior[1].

In a similar way as other forms of collective behavior, synchronization depends on the properties of the oscillators and on the structure of the network of interactions. But it also depends on the initial state of each oscillator. The latter is at the heart of the study of Martens *et al.* They consider certain initial conditions that lead to the coexistence of a synchronized and a desynchronized region in arrays of oscillators: a chimera state[2].

Chimera states are spatiotemporal patterns in which mutually synchronized oscillators, characterized by having identical frequency and hence constant phase differences, coexist with desynchronized ones, which run at different (and not necessarily constant) frequencies. Such states were first reported to exist eight years ago by Kuramoto and Battogtokh[3]. They are generally observed in systems with finite-range nonlocal coupling, in which the oscillators are neither coupled to all the others nor to first neighbors only. All oscillators have the exact same frequency if uncoupled; the coupling is identical for all of them, and potential differences due to boundary effects can be eliminated by considering periodic boundary conditions (e.g., in which the oscillators are organized on a circle or on a torus). Given all the symmetry of the setting it is not intuitive at all that the oscillators could do anything other than to evolve with statistically equivalent oscillatory patterns or provide support for stationary waves. Yet, stable chimeras can exist even when complete synchronization, which is the state that reflects the symmetry of the system, remains stable.

The discovery of chimera states has fundamental implications as it shows that structured dynamical patterns can emerge from otherwise structureless networks. As noted by Abrams *et al.*[4], analogous symmetry breaking is observed in dolphins and other animals that have evolved to sleep with only half of their brain at a time[5]. Neurons exhibit synchronized activity in the sleeping hemisphere and desynchronized activity in the hemisphere that is awake. Moreover, because synchronization is believed to play a central role in information processing (and abnormal synchronization may lead to epilepsy), the extent to which local synchronization is determined by the properties of the underlying network is broadly significant for the study of neuronal networks in general.

Chimera states need not to be frozen—they can propagate as rotating spiral waves. Propagating waves are common in reaction-diffusion systems and have been widely studied in Belousov-Zhabotinsky reaction systems. There too, the pattern will depend on the initial conditions. But there is something that is unique to the states considered by Martens *et al.*: the very fact that they are chimera, i.e., they have not only synchronized but also desynchronized regions. Simulations in two-dimensional arrays of oscillators have shown that the system can self-organize into a desynchronous core surrounded by a spiral wave of synchronized oscillators[6]. Along the arms of the spiral the oscillators have the same phase—they are phase-locked (see Fig. 1). In their study, Martens *et al.*[1] provide the first analytical description that predicts the existence of such spiral wave chimeras. Therefore, as unusual as these spirals may seem, they cannot be attributed to artifacts of the numerical simulations.

Their description is based on considering oscillators of natural frequency $\omega$ whose phase is influenced by the phase of the others through the sine of the phase difference plus a phase lag $\alpha$. The nonlocal coupling is modeled through a Gaussian kernel, which facilitates the use of perturbation theory to predict that, up to order $\alpha$, the (relative) radius of the desynchronized core is $2\alpha/\sqrt{\pi}$ and the angular velocity of the spiral arms is $\omega - \alpha$. This analysis shows that the spiral wave chimeras will exist for small $\alpha$, when both the desynchronized core and the angular velocity of the spiral (in the frame rotating with the natural frequency $\omega$) are small.

These results were obtained for a relatively simple system under analytically treatable conditions. This is, however, yet another reason why they are important, as they reveal what is most likely to be only the tip of the iceberg.

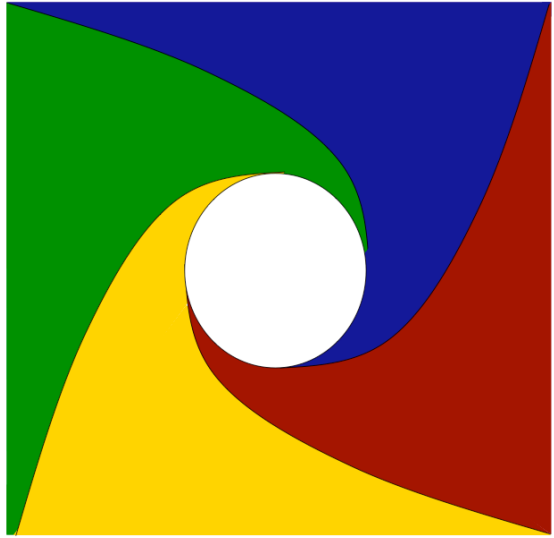

**Figure 1.** Schematic representation of a spiral wave chimera in a two-dimensional array of oscillators with small phase lag. Each color represents a range of phases for the phase-locked oscillators, whereas the white circle corresponds to the desynchronized core.

Open problems abound. The characterization of the basins of attraction associated with chimera states—and the possible attractors, for that matter—remains widely untouched. For instance, how can one determine whether a given set of initial conditions corresponds to a uniform as opposed to a chimera state? The characterization of the stability of these states as functions of the system parameters also remains fairly under-explored. It is possible that for some parameter choices multiple coherent formations will coexist or new forms of non-stationary chimeras will emerge. Experimental observation of such states in natural systems, neural or not, would also be extremely informative. As a matter of fact, although chimera states do not need extra structure to exist, they are not destroyed by small disorder either[7], which certainly strengthens their prospects for real systems. More important, additional structure can lead to a myriad of other possible behaviors, including quasiperiodic chimeras[8] and chimeras that "breathe", in the sense that coherence in the desynchronized population cycles up and down[4].

Future research may benefit from two other surprising recent discoveries. First, for several systems of infinitely many non-identical phase oscillators, it has been shown that a wide class of solutions can be reduced *exactly* (not approximately!) to a system described by just a handful number of degrees of freedom[9]. This has in fact already inspired recent research on chimera states for non-identical oscillators[7]. Second, in complex networks of identical oscillators, it has been demonstrated that the stability of globally synchronous states depends *sensitively* on the structure of the network[10]. It is thus natural to ask about the nature of (partially synchronous) chimera states in such complex networks. If previous experience is anything to go by, one can expect that this research will lead to incongruous yet fascinating new surprises about the dynamics of complex systems.